\documentclass[10pt,journal,compsoc]{IEEEtran}

\ifCLASSOPTIONcompsoc
  \usepackage[nocompress]{cite}
\else
  \usepackage{cite}
\fi
\ifCLASSINFOpdf
  \usepackage[pdftex]{graphicx}
  \graphicspath{{../pdf/}{../jpeg/}}
  \DeclareGraphicsExtensions{.pdf,.jpeg,.png,.jfif}
\else
  \usepackage[dvips]{graphicx}
  \graphicspath{{../eps/}}
  \DeclareGraphicsExtensions{.eps}
\fi
\usepackage{amsmath}
\usepackage{multirow}
\ifCLASSOPTIONcompsoc
  \usepackage[caption=false,font=footnotesize,labelfont=sf,textfont=sf]{subfig}
\else
  \usepackage[caption=false,font=footnotesize]{subfig}
\fi

\usepackage{stfloats}
\usepackage{url}
\newcommand\MYhyperrefoptions{bookmarks=true,bookmarksnumbered=true,
pdfpagemode={UseOutlines},plainpages=false,pdfpagelabels=true,
colorlinks=true,linkcolor={black},citecolor={black},urlcolor={black},
pdftitle={Deep Hierarchical Super Resolution for Scientific Data},
pdfsubject={Machine Learning},
pdfauthor={Skylar Wurster},
pdfkeywords={Machine learning, Hierarchical data, Super resolution}}
\ifCLASSINFOpdf
\usepackage[\MYhyperrefoptions,pdftex]{hyperref}
\else
\usepackage[\MYhyperrefoptions,breaklinks=true,dvips]{hyperref}
\usepackage{breakurl}
\fi

\usepackage[normalem]{ulem}

\usepackage{amsthm}
\usepackage{amssymb}
\usepackage{algorithm}
\usepackage{algpseudocode}
\usepackage[shortlabels]{enumitem}
\usepackage{xcolor}
\newcommand{\comment}[1]{}

\begin{document}
%
\title{Deep Hierarchical Super Resolution \\ for Scientific Data}
%
%
%
%
\author{Skylar~W.~Wurster,
        Hanqi~Guo,~\IEEEmembership{Member,~IEEE,}
        Han-Wei~Shen,
        Tom~Peterka,~\IEEEmembership{Member,~IEEE,}
        and Jiayi~Xu%

\IEEEcompsocitemizethanks{
    \IEEEcompsocthanksitem S. Wurster, H. Guo, H.W. Shen, and J. Xu are with the Department of Computer Science and Engineering, The Ohio State University, Columbus, OH, 43210, USA.\protect\\Emails: \{wurster.18, guo.2154, shen.94, xu.2205\}@osu.edu
    \IEEEcompsocthanksitem T. Peterka is with the Mathematics and Computer Science Division, Argonne National Laboratory, Lemont, IL 60439, USA.\protect\\E-mail: tpeterka@anl.gov
}


}

%
%

\ifCLASSOPTIONpeerreview
\fi
%



\IEEEtitleabstractindextext{%
\begin{abstract}
We present a \textcolor{black}{novel technique} for hierarchical super resolution (SR) with neural networks (NNs), which upscales volumetric data represented with an octree data structure to 
\textcolor{black}{a high-resolution uniform grid}
with minimal seam artifacts on octree node boundaries. 
\textcolor{black}{Our method uses existing state-of-the-art SR models and adds flexibility to upscale input data with varying levels of detail across the domain, instead of only uniform grid data that are supported in previous approaches.}
The key is to use a hierarchy of SR NNs, each trained to perform $2\times$ SR between two levels of detail, with a hierarchical SR algorithm that \textcolor{black}{minimizes seam artifacts by starting from the coarsest level of detail and working up.}
We show that our hierarchical approach
\textcolor{black}{outperforms baseline interpolation and hierarchical upscaling methods, and demonstrate the usefulness of our proposed approach across three use cases including data reduction using hierarchical downsampling+SR instead of uniform downsampling+SR, computation savings for hierarchical finite-time Lyapunov exponent field calculation, and super-resolving low-resolution simulation results for a high-resolution approximation visualization.}
\end{abstract}

\ifCLASSOPTIONpeerreview
\fi
}

\maketitle

\IEEEdisplaynontitleabstractindextext

%
\IEEEpeerreviewmaketitle

\ifCLASSOPTIONcompsoc
\IEEEraisesectionheading{\section{Introduction}\label{sec:introduction}}
\else
\section{Introduction}
\label{sec:introduction}
\fi

%
%
%
%


 

\IEEEPARstart{C}{omputation} \textcolor{black}{resources such as node-hours, storage space, memory, and bandwidth are often limited in supply for scientific computing, which pushes scientists and researchers to develop new strategies to perform the desired tasks quicker and use a smaller storage footprint.
Recently, approaches \cite{volume_upscaling, SSRTVD, SSRVFD, TSRTVD, STNet, fukami_spatial, fukami_spatiotemporal},  have used machine learning (ML) based strategies to reduce computation resources using super resolution (SR) with trained neural networks (NNs).
SR approaches use a curated set of training data to teach a neural network to upscale low-resolution (LR) volumes to their high-resolution (HR) ground truth, and may operate in either the spatial domain, temporal domain, or both \cite{SRCNN, ESRGAN, SSIM, volume_upscaling, SSRTVD}.
With these trained networks, time and computation resources can be saved by running simulations at a lower spatial resolution \cite{SSRVFD, SSRTVD, fukami_spatial} or temporal resolution \cite{TSRTVD, fukami_spatiotemporal, STNet}, and performing SR to infer the missing spatial and/or temporal details by taking advantage of the transformations provided by trained neural networks.
The trained neural network can also be used for data reduction by saving LR data and restoring the HR data via the neural network's upscaling abilities \cite{STNet}.}

\textcolor{black}{The goal of this work is to use SR techniques with hierarchical data representations for increased computation resource efficiency. 
Hierarchical data structures are another category of strategies designed for computation resource efficiency, \textcolor{black}{yet there are no SR works that incorporate hierarchical methods.}
Hierarchical methods, such as octrees, k-d trees, and adaptive mesh refinement (AMR), use an adaptive resolution throughout the spatial domain, which will be more fine (higher resolution) in regions of interest, and more coarse (lower resolution) elsewhere.
These methods have been used to speed up simulations \cite{nyx, AMReX, chombo, LAVA, Enzo}, improve isosurface extraction and volume rendering speed \cite{octree1, octree2, octree3, octree4}, speed up finite time Lyapunov exponent (FTLE) field computation \cite{sadlo07_AMRFTLE, barakat12_adaptiveFTLE}, improve implicit neural network training \cite{acorn}, and assist data reduction tools \cite{precision_resolution_tree}.}

\textcolor{black}{We propose a hierarchical SR method that combines the predictive ability of SR NNs with the resource efficiency of hierarchical methods to improve the performance of SR models across resource-constrained use cases.
There are two primary challenges when incorporating adaptive hierarchical data with neural-network based SR.
First, conventional SR neural networks assume the input will be a regular grid, which is incompatible with hierarchical data formats.
Second, hierarchical methods often display \textit{seams} between blocks of different levels of detail, which can cause distracting visual artifacts in tasks such as volume rendering or isosurface extraction \cite{AMR_interpolation1, AMR_interpolation2, AMR_interpolation3}.}

\textcolor{black}{We design a hierarchical SR algorithm that is compatible with existing regular grid SR models, allowing any state-of-the-art SR model to be used, such as ESRGAN \cite{ESRGAN}, SSRTVD \cite{SSRTVD}, and STNet \cite{STNet}.
Our hierarchical SR algorithm relies on a NN hierarchy, where each network in the hierarchy is responsible upscaling one level of detail to the next more fine level.
To reduce the impact of seams, we first downscale the hierarchical data to a regular grid at the coarsest level, and then use the hierarchy of networks to upscale the coarse regular grid one level of detail at a time.
As the data are upscaled level by level, any higher resolution voxels that are available from the hierarchical data are used to overwrite the upscaled approximated voxels.
This upscaling-overwriting process is repeated until the data is at the desired resolution.
Seams are avoided by performing regular grid super resolution on the entire domain, as opposed to upscaling separate chunks of data with varying levels of detail separately.
To accompany our algorithm, we design a data format called an ``SR-octree'' to hold the hierarchical data in a format that is efficient for our hierarchical SR algorithm by keeping large chunks of adjacent data joined in a single node when possible.}

\textcolor{black}{In our experiments, we use three different architectures (ESRGAN \cite{ESRGAN}, SSRTVD \cite{SSRTVD}, and STNet \cite{STNet}) within our hierarchies, and show they achieve better peak signal-to-noise (PSNR) ratio and SSIM \cite{SSIM} than bilinear/trilinear interpolation.
We verify our hierarchical SR algorithm outperforms a baseline block-wise upscaling algorithm, and effectively minimizes the presence of seams between blocks with different refinement levels.}

\textcolor{black}{We also demonstrate our technique's usefulness for three resource-constrained use cases. 
The first use case is data reduction.
As compute power grows, simulations are run at larger resolutions.
However, storage and bandwidth cannot keep up, creating a bottleneck that leads to the need for data reduction \cite{compression_survey}. 
SR algorithms are one proposed solution to reduce storage overhead by downscaling and using SR to recover details \cite{STNet, SSRVFD, SSRTVD}.
We show that our method applied to a state-of-the-art network STNet \cite{STNet} improves data reduction capabilities.
Our second use case is computational efficiency for hierarchical computation. 
Some expensive scientific and/or visualization algorithms will use hierarchical computation to focus computing resources on regions of interest.
If those are allowed to run with an increased tolerated error, hierarchical super resolution can be used to upscale the results for saved computation time.
In our evaluation, we use FTLE as the specific use case \cite{sadlo07_AMRFTLE, barakat12_adaptiveFTLE}.
Our last use case is upscaling low resolution simulation data for time savings.
High resolution simulations take longer to compute compared to the same simulation at a lower resolution.
Like other SR techniques, our method can be used to upscale low-resolution simulation output to preview what the high resolution simulation may look like. 
In our evaluation, we use the Nyx cosmological simulation \cite{nyx} as the simulation we train our hierarchy to upscale.
}

\textcolor{black}{In summary, our contributions are the following:
\begin{enumerate}
    \item A hierarchical SR algorithm that upscales hierarchical data while minimizing seam artifacts for better reconstruction and less visual artifacts
    \item A hierarchy of SR neural networks for use with our hierarchical SR algorithm
    \item A hierarchical data structure called an SR-octree, which stores hierarchical data for use with our hierarchical SR algorithm.
\end{enumerate}}

\section{Related Works}
We review related works in super resolution methods for images and scientific data and hierarchical methods for scientific computing.  

\subsection{Super resolution}
Super resolution techniques are used to increase the resolution of an image\cite{ESRGAN, LAPSRN}, video \cite{DenseBlockVideoSR}, volume \cite{SRCNN, SSRTVD}, or time-varying volumetric data \cite{TSRTVD}.
The related works below do not include image SR unless otherwise noted, and we refer readers instead to a survey by Wang et al. \cite{ISR_survey} for a comprehensive review on image SR.
In general, there are two categories of SR: spatial super resolution (SSR) and temporal super resolution (TSR). 
TSR increases the temporal resolution of input data, such as TSR-TVD by Han et al. \cite{TSRTVD}, which uses a convolutional LSTM to learn the recurrence between timesteps and recovers interpolated timesteps with higher accuracy than linear interpolation. 
TSR is also useful for video frame interpolation \cite{DenseBlockVideoSR}.

Our approach is an SSR technique, which increases the spatial resolution of input data. 
Convolutional neural networks (CNNs) were adopted for image SR, with specific techniques introduced to improve quality such as residual learning \cite{DenseBlockVideoSR, residual_learning}, adversarial training \cite{SRGAN, ESRGAN}, and removing bias from interpolation techniques used in the networks \cite{LAPSRN, pixel_shuffle}. 
Similar to our approach, LapSRN \cite{LAPSRN} estimates multi-level upscaled output to allow $2\times, 4\times, 8\times, ...$ scale factor SR. 
However, their network trains end-to-end using a Charbonnier loss function and assumes the input is at the lowest resolution, whereas our hierarchy trains each model individually which allows SR from any starting resolution below the training data resolution. 
Weiss et al. use an SR network to increase the resolution of isosurfaces by upscaling the depth and normal maps \cite{isosurface_SR}. 
Weiss et al. also use an SR approach for adaptive sampling SR for isosurface and volume rendering images \cite{adaptiveSR}.

Image-focused models were adapted to accommodate SSR of 3D scientific data as well, but no methods have yet been proposed for hierarchical data SSR.
Zhou et al. \cite{volume_upscaling} use a 3D convolutional neural network (CNN) to perform SSR on volumes with better feature reconstruction than trilinear interpolation or cubic-spline interpolation. 
Guo et al. \cite{SSRVFD} use three NNs to perform SSR on 3D vector fields by training a single network for each variable. 
Xie et al. create tempoGAN \cite{tempogan}, which upscales fluid flows for temporally consistent HR output. 
Fukami et al. compare two ML-based SSR methods for 2D fluid flow \cite{fukami_spatial}. 
Höhlein et al. \cite{weather_SR2} use CNNs to perform wind field downscaling, which learns the mapping from coarse input data to fine predicted output data. 
SSRTVD by Han and Wang \cite{SSRTVD} uses a generative adversarial network (GAN) for upscaling volumes such that the upscaled output frame is temporally coherent with adjacent timesteps. 
With the recent advances in SR, Jakob et al. have created a public 2D flow dataset with varying Reynolds number, citing that ML methods are a powerful method for interpolating flow maps and data reduction \cite{ETH_ZurichDataset}.

There are also works that use ML for spatiotemporal SR, wherein both space and time dimensions are upscaled \cite{fukami_spatiotemporal, STNet}.
These networks use a combination of convolutional layers for spatial upscaling, and may use convolutional and/or recurrent layers for temporal upscaling.

\subsection{Hierarchical data structures}
Hierarchical data structures have been used in applications including rendering, scientific simulations, data reduction, and ML, to reduce computation and storage requirements by more efficiently allocating resources available. 
A survey of octree-based rendering methods is available by Knoll \cite{octree3}.
Rendering pipelines have utilized octrees \cite{octree} for efficient and scalable volume rendering by dynamically loading data from different octree depths depending on the current camera position and angle. 
Gobbetti et al. \cite{octree_rendering1} organize large scalar fields into an octree structure and only feed the GPU information relevant to the current viewpoint and transfer function at each frame. 
Knoll et al. \cite{octree1} create a technique for isosurface ray tracing for large octree volumes. 
Wilhelms and Van Gelder \cite{octree2} use octrees for isosurface generation when the regions of interest are unknown. 
\textcolor{black}{Hadwiger et al. \cite{hadwiger12_microscopy} enable visualization of petascale volumes of microscopy data using an out-of-core multiresolution approach.
Fogal et al. \cite{fogal13_multigridrendering} use an out-of-core adaptive multi-grid rendering algorithm that samples densely only where needed to enable rendering of large scientific data on consumer graphics cards.}
A key problem for rendering hierarchical data is the seam artifact that can occur between data blocks. 
Methods for reducing the effect of this issue are presented by Santos et al. \cite{AMR_interpolation1}, Wald et al. \cite{AMR_interpolation2}, and Wang et al. \cite{AMR_interpolation3}. 

Aside from rendering, hierarchical data formats have also been used during simulation to improve performance. 
Losasso et al. \cite{octree5} create a method for simulating water and smoke on an unrestricted octree to capture small scale visual detail and allow for efficient solving of the Poisson equation. 
Popinet \cite{octree6} uses quadtrees and octrees for a flexible and efficient approach to solving time-dependent incompressible Euler equations. 
Adaptive mesh refinement (AMR) schemes, a similar hierarchical data representation, have been utilized in numerous scientific software packages to improve performance by using a coarse mesh where there is little detail and fine meshes where precision is more important. 
Berger and Colella first design the AMR scheme for use with hyperbolic partial differential equations \cite{AMR2} and two dimensional shock dynamics \cite{AMR1}. 
Different implementations of AMR are implemented in modern simulation packages such as Enzo\cite{Enzo}, Chombo\cite{chombo}, LAVA\cite{LAVA}, Nyx \cite{nyx}, and AMReX\cite{AMReX}. 

Hierarchical representations can also be used for data reduction. Knoll et al. \cite{octree1} use an octree representation that only uses a fraction of the original data's memory footprint, similar to the approach of Velasco and Torres \cite{octree_compression}, to accelerate volume rendering. 
Bhatia et al. \cite{AMM} introduce AMM, an adaptive multilinear mesh framework that reduces the memory footprint of raw data using tensor products of linear B-spline wavelets and allow a tradeoff between numerical precision and spatial resolution. Hoang et al. \cite{precision_resolution_tree} use a precision-resolution tree, and an encoding of the tree, to perform data reduction on scalar field data. 
Ainsworth et al. \cite{scientific_compression} create a multilevel technique to compress univariate data that allows users to select a level that minimizes memory use while meeting some tolerance level. 
Another multilevel technique is presented in MGARD \cite{MGARD}, which uses adaptive error based coefficient quantization to enable different tolerances at different levels of detail, and proposes to use the multilevel data decomposition as a preconditioner to terminate at an appropriate level.
 
Hierarchical data representations have also been used in ML.
However, our approach differs as it is a SR approach \textit{for} hierarchical data representing scalar fields. 
Riegler et al. \cite{OctNet} create octree convolution, pooling, and upscaling layers for a NN to classify occupancy octree data representing geometric objects. 
This method is limited to only occupancy octrees, and not octrees for scalar fields, which we use. 
Tatarchenko et al. \cite{OctreeGeneratingNetworks} create a deep convolutional decoder that learns to decode dense input to octree representations of 3D objects, which faces the same limitation as occupancy octrees. 
Martel et al. \cite{acorn} and Takikawa et al. \cite{implicit_hierarchy} use hierarchical data representations to assist with implicit neural representation.

\section{Approach}

There are three components to our approach - the NN hierarchy, hierarchical data structure, and hierarchical SR algorithm. 
Outlined in \autoref{nnsection} is the construction and training of a SR hierarchy, which is composed of multiple neural networks trained individually.
Next, we discuss the hierarchical data structure we use called an SR-octree in \autoref{dataFormatSection}. 
This data structure is a hierarchical format equivalent to an octree \cite{octree}, but constructed for our hierarchical SR method. 
We introduce a baseline approach for hierarchical SR in \autoref{baseline_section}, and discuss our method for upscaling the hierarchical data with minimal seam artifacts in \autoref{superResolving}.

\subsection{Spatial super resolution hierarchy}
\label{nnsection}

\begin{figure}[ht]
\centering
\includegraphics[width=\linewidth]{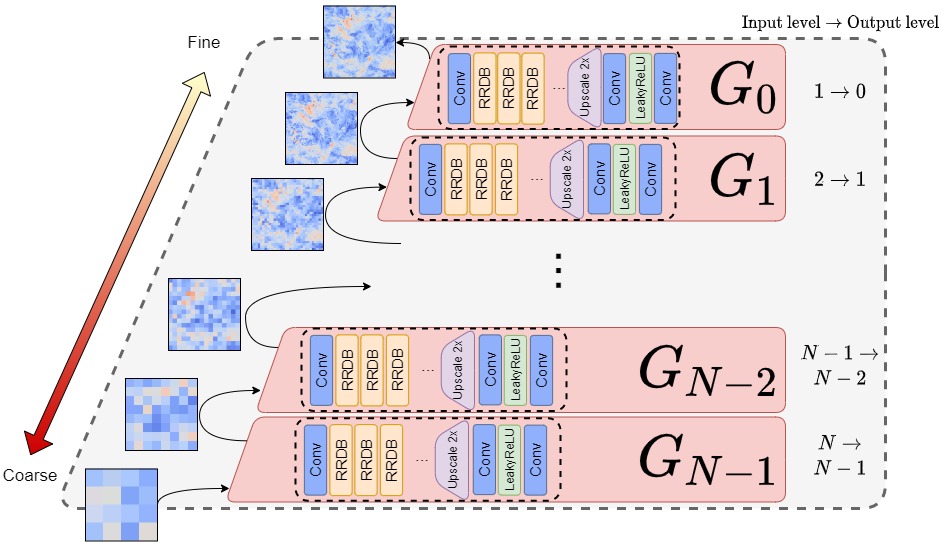}
\centering
\caption{An example of our SR hierarchy composed of $N$ NNs, each trained to perform $2\times$ SR between different downscaling levels. In this example, each $G_i$ is the generator from ESRGAN, but it can actually be any architecture capable of $2\times$ SR.}
\label{fig:hierarchy}
\end{figure}

In this section, we detail the construction of an SR NN hierarchy. 
Since the hierarchical data may be at varying levels of detail, we require a design that can upscale various scales of data.
For example, some regions may need $2\times$ SR, and others may need $16\times$.
Hierarchical NN approaches, such as LapSRN \cite{LAPSRN} or EDSR \cite{EDSR} may offer multiple scale factors, but expect specific input/output levels of detail, which is incompatible with our SR method described in \autoref{superResolving}.

We design a hierarchy compatible with our hierarchical SR approach that is used for uniform scale factor SR at varying scale factors such as $2\times, 4\times, ...$, scaling with the number of NNs in the hierarchy.
Our design works with our hierarchical SR algorithm outlined in \autoref{superResolving} by allowing arbitrary input and output level of detail.
Our design creates a hierarchy of SR networks that allows a volume of any level of detail as input, and produces an upscaled volume of any level of detail between the input and the original HR as output. 
Instead of level of detail, we define ``downscaling level'', such that downscaling level $i$ corresponds to raw data that has been downscaled by a factor of $2^i$, or simply data that is a factor of $2^i$ more coarse than the finest data in the volume. 
As the downscaling level increases, the data become coarser.
We create a hierarchy $G$ of $N$ SR networks $G = \{G_0, G_1, ..., G_{N-1}\}$, where each network $G_i$ is trained to perform $2\times$ SR from downscaling level $i+1$ to $i$. 
By using multiple networks in tandem, larger scale factors like $4\times, 8\times, ...$ up to $2^N\times$ are possible. 
An example of an SR hierarchy is shown in \autoref{fig:hierarchy}, which shows how coarse input is super-resolved as we move up the hierarchy to the finer scales.
Any SR architecture capable of $2\times$ SR can be used in our defined hierarchy for hierarchical SR, and every network within $G$ is identical in terms of architecture, but no weights are shared between models. 
In this paper, we evaluate our hierarchy with ESRGAN \cite{ESRGAN}, SSRTVD \cite{SSRTVD}, and STNet \cite{STNet}.

\subsection{Hierarchical data structure for super resolution}
\label{dataFormatSection}

To model the hierarchical data, we define an octree data structure called an SR-octree. 
Our SR-octree is designed to minimize the number of nodes in the tree and keep data with the same downscaling level contiguous until spatially divided. 
Though everything will be defined in three spatial dimensions (for an octree), the same methods apply to 2D (for a quadtree).
\textcolor{black}{For brevity, we will use the terms ``octree'' and ``SR-octree'' regardless of the dimensionality of the data.}

Like most octrees, children of a parent node in an SR-octree represent the 8 octants that compose the parent. 
The octree data structure only defines the spatial partitioning and \textit{not} spatial averages, where the root node refers to the entire domain. 
Unique to the SR-octree is a $\texttt{dwnscl\_lvl}$ attribute on each node that represents the factor that data have been downscaled in each spatial dimension within that node. 
This attribute allows us to represent contiguous large regions with a uniform downscaling ratio with one node. 
Specifically, each leaf node in the SR-octree holds a volume of data defined by the node's spatial partition (position and size) and downscaling level. 
This is necessary to upscale the data properly in our hierarchical SR algorithm. 

\begin{figure}[ht]
\centering
\includegraphics[width=\linewidth]{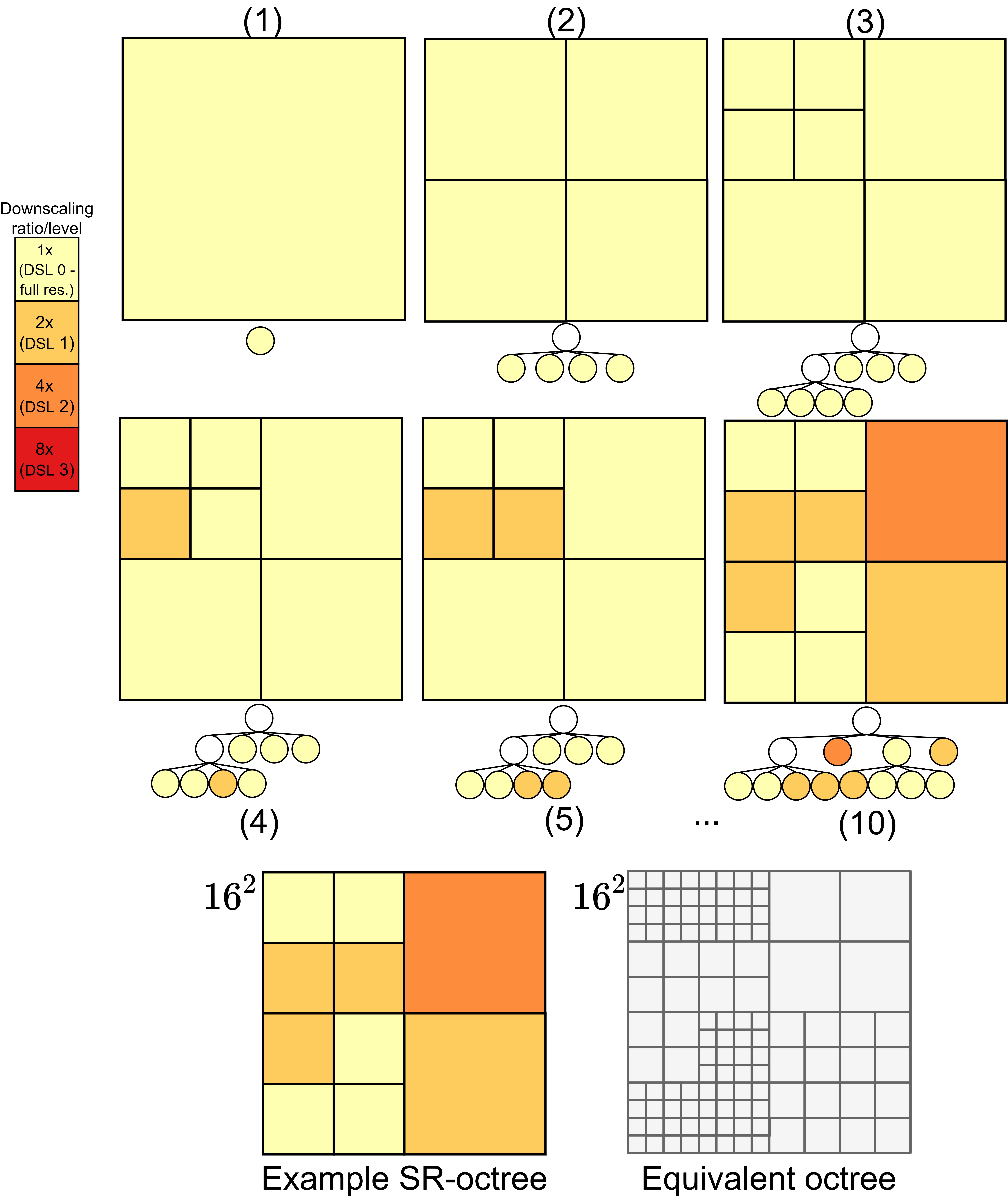}
\centering
\caption{A \textcolor{black}{quadtree} example of an SR-octree and visualization of each quadrant's $\texttt{dwnscl\_lvl}$s while creating hierarchical data for experiments. In (1) data starts at full resolution ($\texttt{dwnscl\_lvl}=0$). In (2), a trial and error step for the $\epsilon$ error is not met, so the volume is split into quadrants, and the process is repeated starting in the top left corner in (3), which also fails, resulting in a second split for that quadrant. In (4), the bottom left quadrant was able to be downscaled while having error less than $\epsilon$, shown by the color changing to orange, which represents $\texttt{dwnscl\_lvl}=1$. In (5), the same downscaling is successful for the sibling of the previous quadrant. By (10), the top right quadrant has been downscaled twice, and the bottom right quadrant once. The bottom shows the equivalent octree \cite{octree} (without the downscaling level attribute) which has a finer mesh over the regions which have a finer resolution.}
\label{fig:octreeHierarchy}
\end{figure}

\autoref{fig:octreeHierarchy} \textcolor{black}{shows the concept of an SR-octree using a quadtree example in 2-D}. 
Non-leaf nodes hold no data and only point to their child octants. 
Nodes are colored according to their downscaling level shown by the key on the left. 
We also show an equivalent ``typical'' octree (as defined by Meagher \cite{octree}) in the bottom right that has more dense nodes in the finer regions.
Notice that the mesh density corresponds to the color (downscaling level) in the equivalent SR-octree on the left compared with the typical octree. 
Our SR-octree construction groups together octree nodes at the same downscaling level into a single node.

\textcolor{black}{In some use cases and for testing purposes, it is necessary to downscale a uniform grid HR volume to an SR-octree, such as for our uses cases for data reduction (\autoref{data_reduction}) and hierarchical SR for FTLE data (\autoref{hierarchical_FTLE})}.
To create an SR-octree from a uniform full resolution volume, we hierarchically reduce a full-resolution volume according to a user set error bound $\epsilon$, as follows.
Starting with the original volume, downscale the volume by $2\times$ in each spatial dimension, and see if the $L^{\infty}$ norm of the downscaled version compared with the original volume is below the threshold $\epsilon$.
If so, keep the volume downscaled by $2\times$ and increment the downscaling level of the volume's octree node by 1.
Otherwise, return the volume to the higher resolution version, split the volume into its octants, and repeat the above process for each suboctant if each dimension of the suboctant is larger than some minimum size $\texttt{min\_chunk}$.
This process will maximally hierarchically downscale the volume according to the error $\epsilon$.
We add other downscaling parameters $\texttt{max\_dwnscl\_lvl}$, which controls the maximum downscaling factor, and 
$\texttt{min\_dwnscl\_lvl}$, which will automatically downscale the original volume to that scale factor before performing the hierarchical downscaling.
After reduction, we join any adjacent nodes that have the same downscaling level so that the hierarchical data is represented with the fewest number of nodes as possible.

The resulting SR-octree has volume data in leaf nodes, each having an associated $\texttt{dwnscl\_lvl}$ attribute for the volume it represents. 
This allows us to easily determine what SR scale factor is needed to restore an octree node to its original full resolution. We note that in visualizations of an SR-octree, the coarseness/fineness of a region is not determined by the density of the grid in a region (such as with typical octrees), but instead is represented with the $\texttt{dwnscl\_lvl}$ attribute, which we visualize with a discrete color scale in \autoref{fig:octreeHierarchy}.

\begin{figure*}[ht]
\centering
\includegraphics[width=\textwidth]{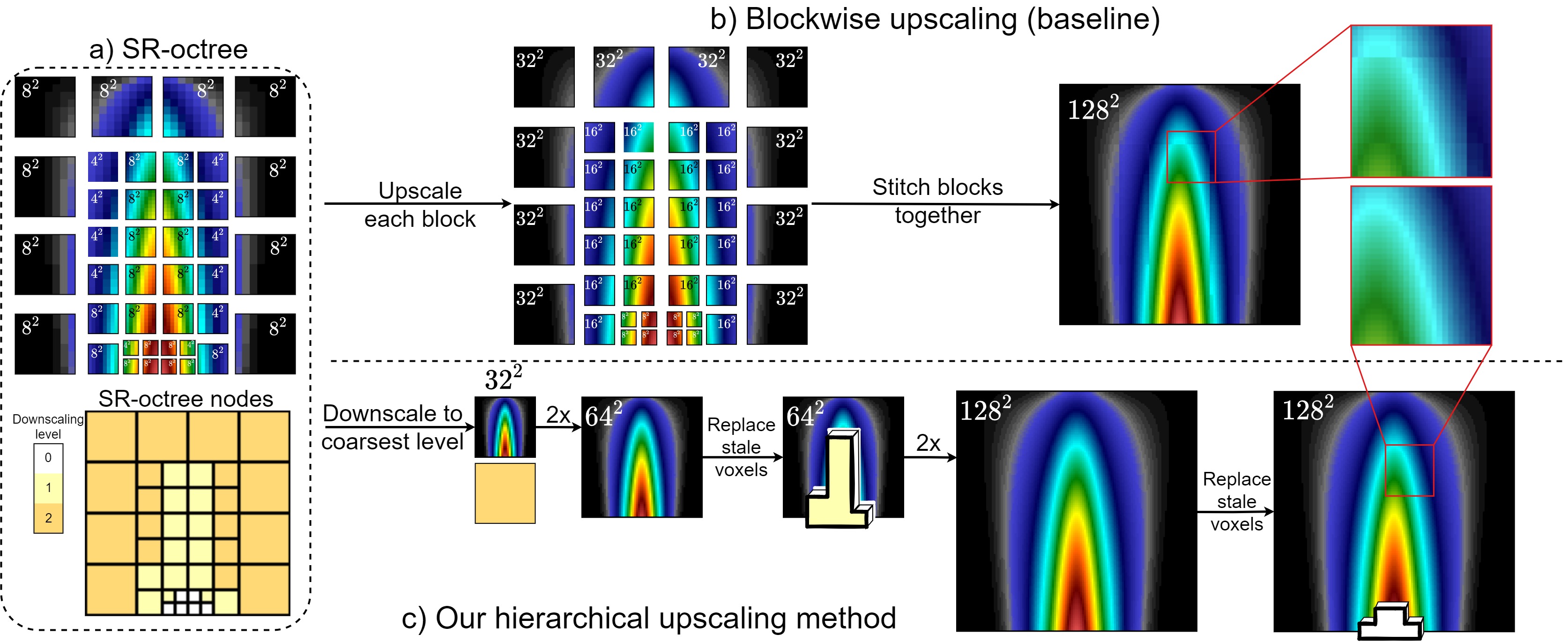}
\centering
\caption{For an SR octree (a), visualization of the baseline blockwise upscaling approach (b) and our hierarchical SR approach (c). In the baseline approach (b), each octree node is upscaled individually and stitched together. In our hierarchical SR approach (c), we follow the approach outlined in \autoref{superResolving} that downscales data to the coarsest level first, then iteratively upscales with networks from the NN hierarchy (d) and overwrites stale voxels.}
\label{fig:BlockwiseComparison}
\end{figure*}

\subsection{Limitation of an alternative hierarchical super resolution method}
\label{baseline_section}
Since no methods currently exist for using SR NNs for upscaling hierarchical data, we \textcolor{black}{contrast our} algorithm with another baseline approach. 
The baseline solution to upscale an SR-octree upscales each octree node individually based on the \texttt{dwnscl\_lvl} attribute each octree node is saved with (i.e. upscales by $2^{\texttt{dwnscl\_lvl}}\times$), and then stitches together the resulting patches, illustrated in the top portion of \autoref{fig:BlockwiseComparison}.

\textcolor{black}{There are two drawbacks to this method that lead to poor performance with trained neural networks and the introduction of \textit{seam} artifacts.
The first problem is that convolutional neural networks use \textit{padding} within convolution layers, which introduces incorrect neighborhood information to the trained neural network, causing error in the upscaled result.}
In convolutional neural networks, padding is a technique that will \textcolor{black}{surround} an input volume with a layer of \textcolor{black}{values} around the boundaries before a convolution operation. 
In practice, padding is used to ensure that the output resolution of the convolutional layer within the NN is the same as the input resolution. 
Without padding, a convolution layer with a stride of 1 would reduce the spatial resolution by $2 \times \lfloor\frac{\texttt{kernel\_size}}{2}\rfloor$ voxels in each direction. 
The network's output is then incorrectly influenced by the padded
\textcolor{black}{values} that are not actually there. 
When composing multiple convolutional layers in sequence, the effect is increased. 
Due to the error from padding, some evaluations for proposed SR NNs choose to ignore some number of border pixels when calculating the recovered image's PSNR, SSIM, or perceptual quality \cite{SR_borders1, SR_borders2, SR_borders3, SR_borders4, SR_borders5}.

\textcolor{black}{Closely tied to the padding problem is the second problem, which is the lack of shared neighbor information.}
Deep convolutional neural networks will often have large \textit{receptive fields}, which indicate the window size that influences each output voxel. 
The larger the receptive field, the more global information is used within the neural network to determine the estimated output voxel value. 
The receptive field is determined by the neural network architecture (number of convolutions and the stride/kernel size/padding, etc.).
Therefore, a hierarchical upscaling algorithm that shares or copies a static number of neighboring voxels/neighbor octree nodes, such as work by Ljung et al. \cite{AMR_interpolation1} or Wang et al. \cite{AMR_interpolation3}, would only work for networks that have a receptive field smaller than the number of shared voxels.

\textcolor{black}{In our setting specifically, padding and limited shared neighborhood information will have a pronounced impact for SR-octree nodes that have few voxels.
For example, a $2^3$ sized leaf node in an SR-octree would be dwarfed by the total number of padded voxels from up to 30 padding operations in our deep convolutional neural networks.
Additionally, in this baseline approach, each node will use its own padding operations, so the upscaled results between each node will not align well, introducing \textit{seams}} between data chunks which can be distracting and/or misleading in visualizations. 

\subsection{Hierarchical super resolution for SR-octrees}
\label{superResolving}

In this section, we present our approach for hierarchical SR.
We design an approach that uses the entire domain at each downscaling level during SR so that padding does not create seams within the volume, and so that the receptive field of the neural network is not limited when upscaling, both of which are limitations outlined in \autoref{baseline_section}.
Though data are distributed across many different downscaling levels, the only level that can \textit{uniformly} represent the hierarchical data is the coarsest level.
Converting the hierarchical data to this uniform representation is the first part of our algorithm and is called the downscaling process, outlined in \autoref{downscaling}.
Next, our upscaling process (detailed in \autoref{upscaling}) will uniformly super resolve the uniform LR data without introducing seam artifacts and overwrite voxels that have more accurate information in the SR-octree, one level at a time, beginning with the coarsest level.
In the end, we are left with an approximation of the HR uniform volume.

\subsubsection{Downscaling process}
\label{downscaling}

Our downscaling process is a preprocessing step which operates on the SR-octree that downscales each node's data to the maximum downscaling level, which gives our upscaling process a uniform volume to begin SR. 
Downscaling all octree nodes to the maximum downscaling level, $\texttt{MAXDSL}$, and putting them together into a regular grid volume is not always possible however. As an example, a single voxel octree node cannot be downscaled by a factor of 8.

We resolve this issue with an iterative downscaling process.
Instead of taking one large downscaling step for each node, we take many smaller global downscaling steps and combine data between each downscaling operation.
Our SR-octree construction guarantees that single voxel octree nodes at $\texttt{MINDSL}$, the minimum downscaling level present in any octree node of an SR-octree $T$, must have siblings with the exact same size (a single voxel) and downscaling level.
Therefore, these siblings can be joined to make a larger octree node that is $2^D$ and can be downscaled by $2\times$. 

\begin{algorithm}
\caption{Our downscaling procedure that takes an octree $T$ as input and returns a LR regular grid volume.}\label{alg:downscale}
\begin{algorithmic}[1]
\Procedure{hierarchical\_downscale}{$T$}
    \State $\texttt{MINDSL} \gets 0$
    \While{$\texttt{MINDSL} < T.\texttt{MAXDSL}$}
        \State $N = \{n \in T ~|~ n.\texttt{dwnscl\_lvl}=\texttt{MINDSL}\}$
        \State Combine adjacent single voxel data nodes in $N$
        \State Downscale each $n \in N$ by $2\times$ in each dim.
        \State $\texttt{MINDSL} \gets \texttt{MINDSL} + 1$
    \EndWhile
    \State $\textbf{return}~T.\texttt{to\_volume()}$
\EndProcedure
\end{algorithmic}
\end{algorithm}

This motivates the downscaling process followed in Algorithm \ref{alg:downscale}, which operates on an SR-octree $T$. 
At each level starting with the finest level, we combine single voxel nodes, and then downscale all nodes at that level by $2\times$ to the next level. 
The procedure is repeated until reaching $\texttt{MAXDSL}$, which is the maximum downscaling level on any node in $T$.
The effect after finishing is that all data from the octree $T$ is downscaled to a single regular grid volume at $\texttt{MAXDSL}$.

Since we downscale regions of data by a factor of 2 multiple times, we require the following property in the downscaling algorithm used. Given a full resolution volume $V$, scale factor $S$, and downscaling algorithm $\texttt{down}(V, S) : \mathbb{R}^{|V|} \rightarrow \mathbb{R}^{|V|/2}$, it must hold that $\texttt{down}(V, 2 \times S) = \texttt{down}(\texttt{down}(V, S), S)$. When this property holds, there is no effect from downscaling a region multiple times compared to downscaling it once at a large downscaling factor. For instance, downscaling a region by a factor of $2\times$ three times would be the same as downscaling it by $8\times$ once. We observe this property in all pooling algorithms (min, max, mean pooling) as well as subsampling, but not in linear interpolation downscaling.

\subsubsection{Upscaling process}
\label{upscaling}

The upscaling process (illustrated in the bottom of \autoref{fig:BlockwiseComparison}, and defined in Algorithm \ref{alg:upscaling}) begins with the uniform LR volume created by the downscaling process, and iteratively upscales and replaces specific voxels with non-approximated data from the SR-octree until we are left with a uniform HR volume approximating the ground truth data.

\begin{algorithm}
\caption{Our upscaling procedure that takes a LR volume $V$ as input (the result of the downscaling procedure described in \autoref{downscaling}) and the corresponding SR-octree $T$. The procedure returns a HR regular grid volume. }\label{alg:upscaling}
\begin{algorithmic}[1]
\Procedure{hierarchical\_upscale}{$V, T$}
    \State $\texttt{MAXDSL} \gets T.\texttt{MAXDSL}$
    \While{$\texttt{MAXDSL} > 0$}
        \State $V \gets \texttt{upscale}(V)$
        \State $\texttt{MAXDSL} \gets \texttt{MAXDSL} - 1$
        \ForAll{$n \in T ~\textbf{s.t.}~ n.\texttt{dwnscl\_lvl} \leq \texttt{MAXDSL}$}
            \State $V[n.\texttt{region}] \gets n.\texttt{data}$
        \EndFor
    \EndWhile
    \State $\textbf{return}~V$
\EndProcedure
\end{algorithmic}
\end{algorithm}

Beginning with the uniform LR volume (argument $V$ in Algorithm \ref{alg:upscaling}), we uniformly upscale this volume to $\texttt{MAXDSL}-1$ via a $2\times$ upscaling algorithm such as a neural network (line 4 in Algorithm \ref{alg:upscaling}).
We do not need to use approximated voxels when non-approximated data exists in the SR-octree. 
If an octree node in the octree has $\texttt{dwnscl\_lvl}\leq \texttt{MAXDSL}$, we replace the approximated voxel values in our volume with the more accurate voxel values from the octree node representing them (line 7). 
We call these approximated values that should be overwritten \textit{stale voxels}, and highlight these regions with an extrusion in \autoref{fig:BlockwiseComparison}. 
We repeat our upscaling $\rightarrow$ stale voxel overwriting until we reach $\texttt{dwnscl\_lvl}=0$. 
Since we upscale the full volume instead of each octree node individually, we do not introduce seam artifacts on octree node boundaries.
Additionally, since the entire domain is being used during upscaling, we are allowing as large a receptive field as possible to the neural network upscaling it.

\subsection{SR model architecture design}
\label{architectures}

\textcolor{black}{In our proposed NN hierarchy, any spatial SR model capable of $2\times$ upscaling is compatible. 
In this paper, we use ESRGAN \cite{ESRGAN}, SSRTVD \cite{SSRTVD} and STNet \cite{STNet}, which are all state-of-the-art neural networks for SR.
Small changes were made to the models for compatibility and training stability.
For all three models, we find they train more stably and with better reconstruction without the discriminator network(s) or losses, so each of the three architectures are only regressors that upscale input.
Additionally, we adjust the models to only perform $2\times$ upscaling instead of $4\times$ upscaling.
This is done by using only a single pixel/voxel shuffle \cite{pixel_shuffle} at the end of the model to upscale the features by a factor of 2 in each dimension instead of what the original architectures suggests.}
Regardless of the architecture chosen for the NN hierarchy, each generator network $G_i$ in our constructed neural network hierarchy is responsible for learning $2\times$ upscaling on input data of downscaling level $i+1$, generating downscaling level $i$.
Our loss function for \textcolor{black}{any network} during training is the L1 loss $\mathcal{L}_{1}(V, V^{GT})$, where $V$ is the upscaled result from the model, and $V^{GT}$ is the ground truth data.

\section{Evaluation}
\label{results}

We evaluate our method on
\textcolor{black}{seven} datasets, described in \autoref{datasets}. 
In \autoref{nntraining}, we describe hyperparameters and the training procedure. 
\textcolor{black}{In \autoref{baseline_comparison}, we compare our method against baselines for both uniform and hierarchical SR}. 
In \autoref{blockwise}, \textcolor{black}{we show that our hierarchical SR approach out-performs a baseline block-wise upscaling approach for hierarchical data, and that our upscaled volumes minimize seam artifacts}. 
\textcolor{black}{In \autoref{use_cases}, we apply our method for three use cases spanning data reduction, compute time reduction by upscaling adaptive resolution algorithm results, and compute resource reduction by upscaling low resolution simulations for a visualization preview of the high resolution result}.
All PSNR/SSIM metrics listed are calculated in the data space as opposed to image space.
\textcolor{black}{Additionally, whenever bilinear or trilinear interpolation is used, the values are assumed to be at cell centers instead of cell corners (\texttt{align\_corners=False}), which consistently provides more accurate reconstruction.}

\begin{figure*}[ht]
\centering
\includegraphics[width=\textwidth]{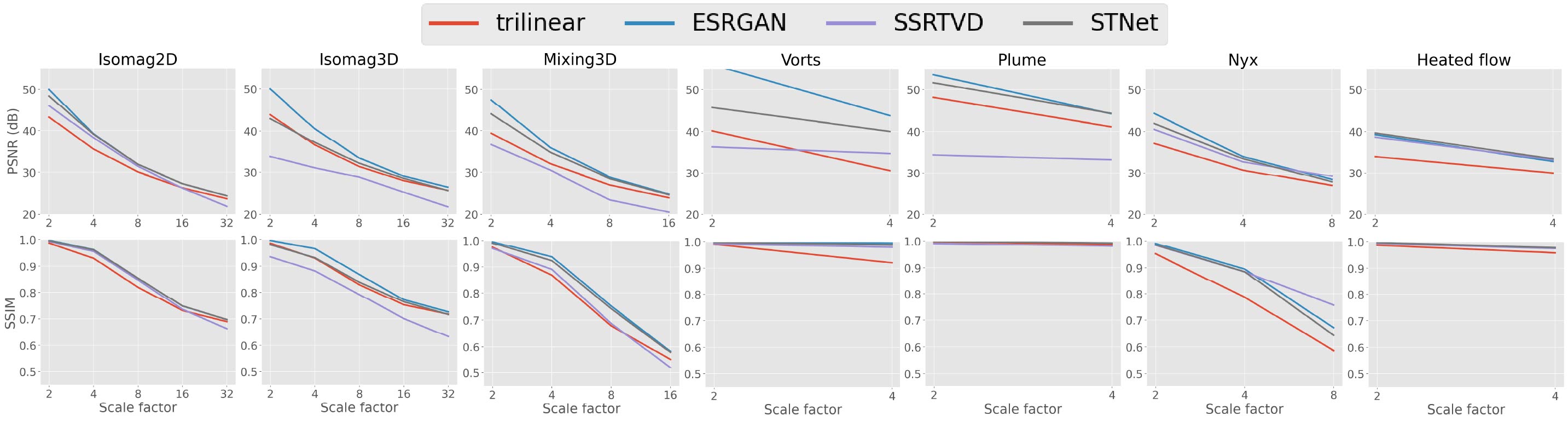}
\centering
\caption{Uniform SR PSNR (dB, top row) and SSIM (bottom row) for hierarchies trained with ESRGAN (blue), STNet (purple), and SSRTVD (gray) compared to bilinear/trilinear interpolation (red). The x-axis represents the scale factor, with scale factors larger than $2\times$ tested using multiple networks from the trained hierarchy in tandem.}
\label{fig:uniformSRquant}
\end{figure*}

\subsection{Datasets}
\label{datasets}
We experiment with seven scalar field datasets. 
Three are from Johns Hopkins Turbulence Databases (JHTDB) \cite{JHUTDB1} that are time-varying results from a direct numerical simulation (DNS) for turbulent fluid flow. 
Information about each dataset is shown in \autoref{datasets_table}. 
\textbf{Isotropic3D} is a 3D velocity magnitude dataset from a DNS of isotropic turbulent fluid flow at Reynolds number $R_{\lambda}\sim 433$, hosted in the ``isotropic1024coarse'' dataset in the JHTDB. 
We take the middle z-axis slice (z=512) of the Isotropic3D magnitude dataset to create an \textbf{isotropic2D} dataset for testing our approach with 2D data. 
The \textbf{mixing3D} dataset is the velocity magnitude field from the ``mixing'' dataset in JHTDB. 
The solar \textbf{plume} dataset is a velocity magnitude field from a simulation that examines the effect of the solar plume on the heat, momentum, and magnetic field of the sun. 
The \textbf{vorts} dataset is a $128^3$ vorticity magnitude scalar field dataset from a pseudo-spectral simulation of vortex structures.
\textcolor{black}{\textbf{Nyx} is a cosmological simulation which is run with parameter settings uniformly sampled within suggested ranges by domain scientists: total matter density $OmM \in [0.12, 0.155]$, total density of baryons $OmB \in [0.0215, 0.0235]$, and Hubble constant $h \in [0.55, 0.85]$ \cite{nyx, InSituNet}.
\textbf{Heated flow} is a 2D fluid simulation from the Gerris flow solver \cite{gerrisflowsolver} created by G{\"u}nther et al. \cite{Guenther17} that simulates flow around a heated cylinder with Boussinesq approximation.
All datasets are linearly scaled to [0.0,1.0] except for Nyx which is log scaled before being scaled to [0.0, 1.0].
All datasets are represented at single precision floating point accuracy to accommodate the architectures used.}

\begin{table}[ht]
\centering
\caption{Dataset names, their sizes, the number of models $|G|$ in a hierarchy $G$ for each dataset, the number of timesteps for training ($N_{\text{train}}$) and testing ($N_{\text{test}}$), and the number of iterations of training per model in the hierarchy.}
\begin{tabular}{l|c|c|c|c|c}
Dataset  & Volume size                & $|G|$ & $N_{\text{train}}$ & $N_{\text{test}}$ & \textcolor{black}{iters.}\\ \hline \hline
Isomag2D & $1024^2$           & 5     & 400         & 100 & 20000    \\
Isomag3D & $1024^3$ & 5     & 40          & 10    & 4000  \\
Mixing3D & $512^3$    & 4     & 80          & 20  & 8000    \\
Vorts    & $128^3$    & 2     & 20          & 9    & 8000   \\
Plume    & $512\times128^2$    & 2     & 20          & 9      & 15000 \\
Nyx    & $256^3$    & 3     & 170          & 100   & 17000   \\
Heated flow    & $144\times448$    & 2     & 7530          & 7530 & 30120    

\end{tabular}
\label{datasets_table}
\end{table}

\subsection{Network training and hyperparameters}
\label{nntraining}
Three NN hierarchies are trained per dataset using the ESRGAN, STNet, and SSRTVD architectures.
To train $G_i$ in a hierarchy $G$, a ground truth volume $V$ is downscaled to LR $V^{\textbf{LR}}$ and HR  $V^{\textbf{HR}}$ (if necessary, i.e. $i$ is not 0) by scale factors $2^{i+1}$ and $2^i$, respectively. 
Then, the loss function is performed on $G_i(V^{\textbf{LR}})$ and $V^{\textbf{HR}}$. 

The number of models trained in each hierarchy is shown in \autoref{datasets_table}, column $|G|$, and the total number of iterations of training (with batch size 1) is listed in column ``iters''. 
We use Python 3.9 with PyTorch to train on an NVidia A100 40GB Tensor Core graphics card. 
The full resolution 3D data cannot fit in the GPU's memory during training, so we crop training volumes to $96^3$ and use a random starting position for the cropping, augmented with random flipping in any subset of the spatial dimensions to further increase training diversity.

\textcolor{black}{The SSRTVD models are constructed as implemented by the paper with no changes to the number of kernels per convolution.
For our ESRGAN and STNet models, all layers (aside from output) use 96 kernels for a good mix between accuracy, GPU memory use, and training time.
All convolution operations in all networks use reflection padding during training and inference.}

\textcolor{black}{The training time per iteration and the storage of each model are listed in \autoref{training_data_table}.
Since each architecture is the same size across all datasets, and the inputs are cropped to a fixed resolution, the models train at a similar time per iteration regardless of which level of detail they are training on.
STNet trains the quickest, as it is the most lightweight architecture.
ESRGAN trains quicker than SSRTVD, yet is a larger model with more parameters, likely due to SSRTVD's decision to be a wider network with more kernels per convolution instead of deeper network, which uses more convolution layers like ESRGAN does.}

\begin{table}[ht]
\centering
\caption{\textcolor{black}{Training time per iteration and storage cost for each the 2D/3D versions of the models tested with.}}
\begin{tabular}{l|c|c}
Model  & $t_{\text{train}}$ per iter. & Storage size per model  \\ \hline \hline 
ESRGAN 2D   &   0.033 sec.  &   11.56 MB   \\
SSRTVD 2D   &   0.041 sec.  &   8.72 MB   \\
STNet 2D    &   0.017 sec.  &   5.89 MB   \\ \hline
ESRGAN 3D   &   0.514 sec.  &   38.42 MB   \\
SSRTVD 3D   &   0.641 sec.  &   26.98 MB   \\
STNet 3D    &   0.451 sec.  &   21.43 MB     
\end{tabular}
\label{training_data_table}
\end{table}

\subsection{Baseline comparison}
\label{baseline_comparison}
We compare our proposed approach in two steps. 
In \autoref{UniformSR}, we isolate the neural network hierarchies by comparing the uniform grid SR performance of the hierarchies against bilinear/trilinear interpolation.
In \autoref{blockwise}, we isolate our hierarchical SR algorithm by comparing the results of hierarchical SR with our method against the baseline blockwise approach.

\subsubsection{Improved SR with trained hierarchy vs. linear interpolation}
\label{UniformSR}
We perform uniform SR using the trained neural network hierarchies and trilinear interpolation on the test sets for each dataset, and report the median PSNR and SSIM in \autoref{fig:uniformSRquant}.
\textcolor{black}{Our results show that the trained hierarchies consistently outperform bilinear/trilinear interpolation across all scale factors for both PSNR and SSIM aside from a few cases with the SSRTVD models specifically.
The low performance of SSRTVD is not due to our method, as even the $2\times$ scale SSRTVD experiments that do not use our neural network hierarchy do not perform well.
More experiments for training routines may improve results for that model specifically.
Aside from those under-performing models, the results verify that using multiple trained NNs in tandem for SR factors of $4\times$ or higher still performs well which is important because the networks were not trained conditional on the output of previous networks. }

\begin{figure*}[ht]
\centering
\includegraphics[width=\linewidth]{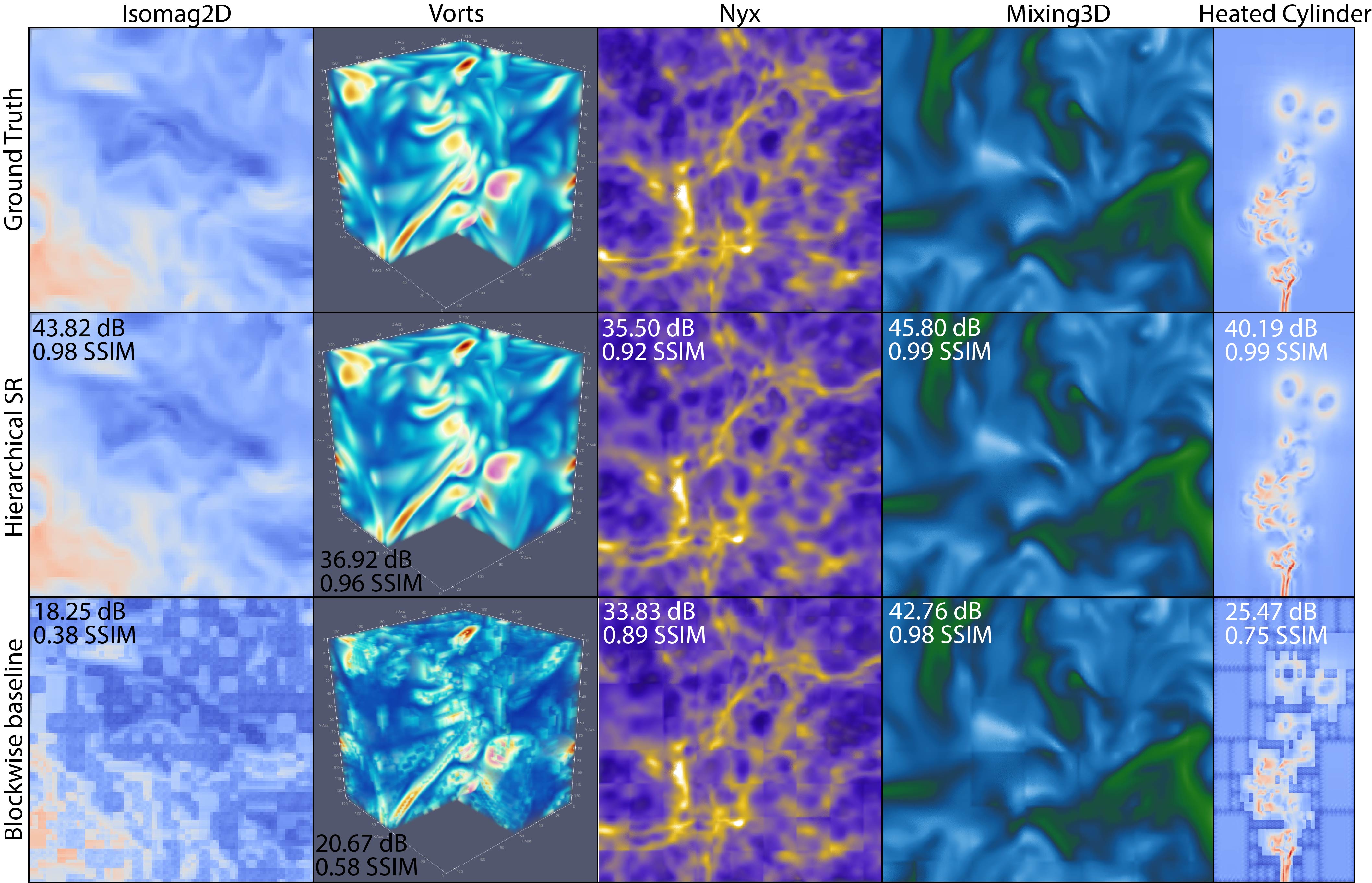}
\centering
\caption{Comparing the baseline blockwise upscaling against our hierarchical SR algorithm with a pre-trained ESRGAN hierarchy. Artifacts are visible in the blockwise baseline, creating seams along block boundaries.}
\label{fig:BlockwiseVsOursESRGAN}
\end{figure*}

\subsubsection{Block-wise baseline vs. hierarchical super resolution}
\label{blockwise}
To evaluate our hierarchical SR algorithm, we generate SR-octrees according to the process described in \autoref{dataFormatSection} and upscale them with the baseline blockwise approach as well as our proposed method \textcolor{black}{using the ESRGAN hierarchies, as they performed best over all datasets.}
\autoref{fig:BlockwiseVsOursESRGAN} shows that blockwise upscaling fails to upscale the octree data properly due to the small size of some octree blocks\textcolor{black}{ which limits the information the network can gather from neighboring voxels, as discussed in \autoref{baseline_section}}.
Our approach does not have this problem because our upscaling process uses the global domain at each step, minimizing the ratio of padding voxels to actual input voxels.

\begin{figure}[ht]
\centering
\includegraphics[width=\linewidth]{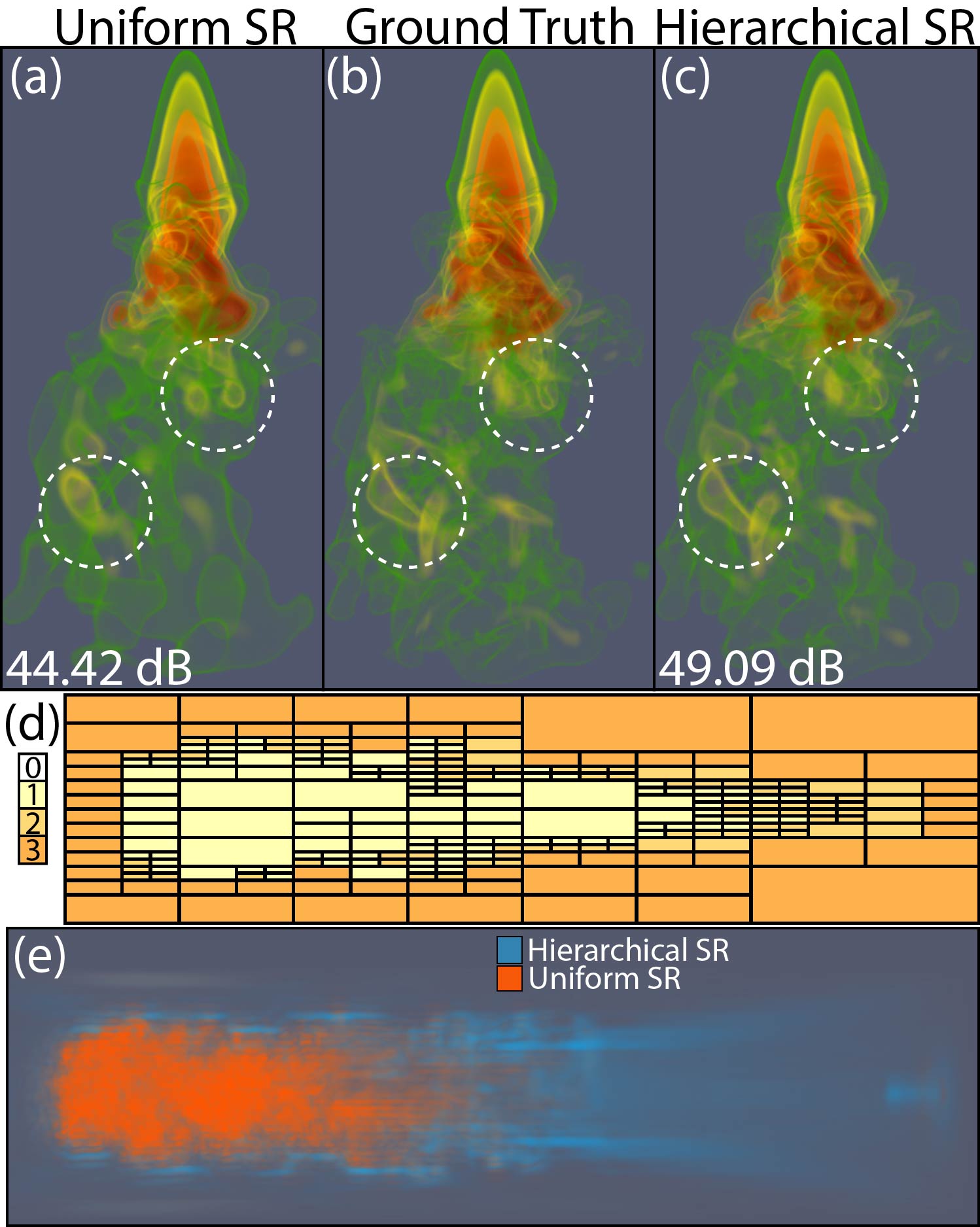}
\centering
\caption{Comparing $4\times$ uniform SR against hierarchical SR at the same data reduction factor ($64\times$). Our approach (c) performs better than uniform SR (a) at the same data reduction factor when comparing to the ground truth (b). The downscaling levels (d) show which regions are more fine (closer to white) and which are more coarse (closer to orange). The error volume (e) shows uniform SR has error near the more turbulent region of interest and our approach has more error in the coarse outer regions where it downscaled more.}
\label{fig:UniformVSmultires}
\end{figure}

\begin{figure}[ht]
\centering
\includegraphics[width=\linewidth]{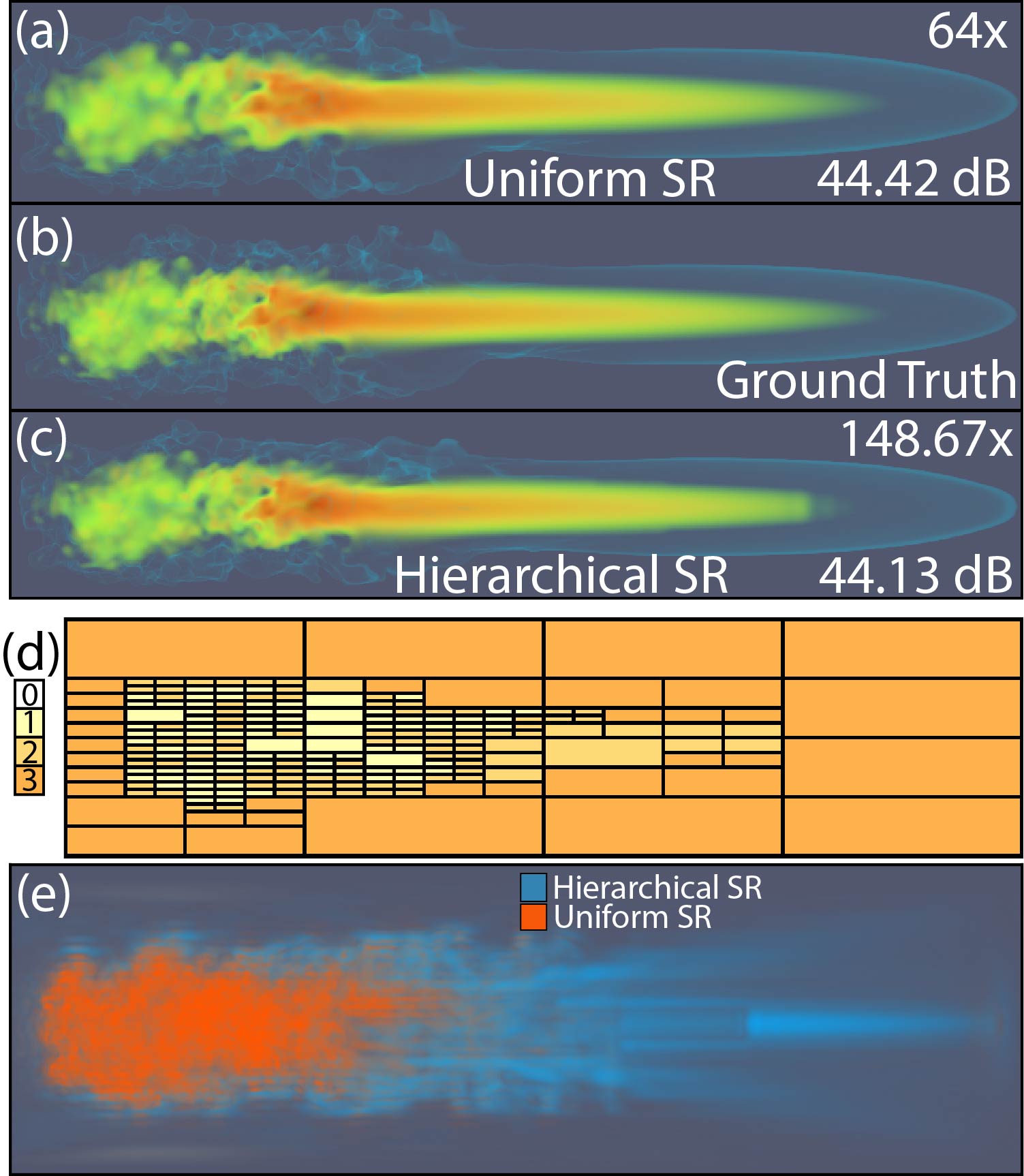}
\centering
\caption{Comparing $4\times$ uniform SR (a) with hierarchical SR (c) when holding the output quality constant. Reduction factors are listed in the top right corner of the image. Both methods achieve just above 44 dB PSNR. Hierarchical SR's reduction rate is $2.32\times$ larger than uniform SR at the same quality. The downscaling levels (d) show which regions are more fine (closer to white) and which are more coarse (closer to orange). The error volume (e) shows that hierarchical SR error is distributed in the coarse outer region of the plume, since it was downscaled further.}
\label{fig:UniformVSmultires_psnr}
\end{figure}

\subsection{Use cases}
\label{use_cases}
\textcolor{black}{We provide examples of three use cases for our hierarchical SR approach.
In \autoref{data_reduction}, we show how our hierarchical SR method can be used to improve STNet's data reduction capabilities by decomposing the spatial domain based on the presence of features instead of uniform downscaling.
\autoref{hierarchical_FTLE} shows how computation resources can be saved on the expensive FTLE field computation by increasing tolerated error in hierarchical FTLE calculation algorithms and using our hierarchical SR algorithm to approximate the HR uniform FTLE.
Lastly, \autoref{nyx_upscaling}, we demonstrate that a trained hierarchy can be used to upscale LR simulation output for visualization, which can be used to save computation resources before running the higher resolution simulation.}

\subsubsection{Data reduction with SR frameworks}
\label{data_reduction}

\textcolor{black}{
As compute power grows, simulations are run at larger resolutions.
However, storage and bandwidth cannot keep up, creating a bottleneck that leads to the need for data reduction \cite{compression_survey}. 
Super resolution methods are one such proposed technique for data reduction \cite{SSRTVD, SSRVFD, STNet}.
One example of an SR approach for data reduction is STNet, which is a state-of-the-art spatiotemporal SR neural network by Han et al. \cite{STNet} that shows promising performance for data reduction compared to TTHRESH \cite{TTHRESH}, a state-of-the-art compressor.
In the data reduction pipeline for STNet, spatiotemporal volumes are uniformly downscaled in both space and time dimensions to a LR before saved to storage.
The data can then be reconstructed via inferring the HR data through the trained STNet model.}

\textcolor{black}{We show that by adding our hierarchical SR capability to STNet's spatial component, the reconstruction quality for the same data reduction factors increases.
In STNet's approach, data is uniformly downscaled by $4\times$ in the spatial domain, which is a data reduction factor of $64\times$ before lossless compression. 
In our approach with a hierarchical STNet, we perform hierarchical downscaling as defined in \autoref{dataFormatSection} with settings $\epsilon=0.02285$, $\texttt{min\_dwnscl\_lvl}=1$, $\texttt{max\_dwnscl\_lvl}=3$, and $\texttt{min\_chunk}=2$, which creates a hierarchical version that has an equivalent data reduction factor (before lossless compression) of $64.03\times$.}

\textcolor{black}{We use the same STNet model trained on the plume dataset which is reported in \autoref{UniformSR}.
In \autoref{fig:UniformVSmultires}, we show the results of using the pre-trained hierarchy for either uniform ($4\times$) SR on the uniformly downscaled data in (a) versus hierarchical upscaling for the SR-octree of the same storage size in (c).
We see that our hierarchical approach improves PSNR by roughly 5 dB \textcolor{black}{for the same data reduction factor}.
We visualize the error of each method in (e), with the uniform upscaling error shown in orange and the hierarchical upscaling error shown in blue.
The uniform upscaling has most of its error in the center more turbulent regions of the volume, whereas the hierarchical upscaling has its error further toward the homogeneous regions that are not of interest.}

\textcolor{black}{For further evaluation, we show that our approach can give higher data reduction at the same target quality.
We identify an SR-octree that causes our hierarchical SR approach to have a similar reconstruction quality to the uniform SR from the last experiment with parameters $\epsilon=0.068$, $\texttt{min\_dwnscl\_lvl}=1$, $\texttt{max\_dwnscl\_lvl}=3$, and $\texttt{min\_chunk}=2$.
Shown in \autoref{fig:UniformVSmultires_psnr}, our hierarchical SR approach was able to achieve 44.13 dB PSNR on this octree, which is roughly the same quality as uniform SR (44.42 dB). 
However, our approach achieves this quality with a data reduction factor of $148.67\times$, which is $2.32\times$ higher than the uniform scale factor data reduction ratio of $64\times$.
The error volume visualization shows a similar trend to the last experiment - uniform SR has a majority of its error (orange) toward the base of the plume with high turbulence, whereas our hierarchical approach has more error (blue) near the coarse regions of the plume. }

\begin{figure}[ht]
\centering
\includegraphics[width=\linewidth]{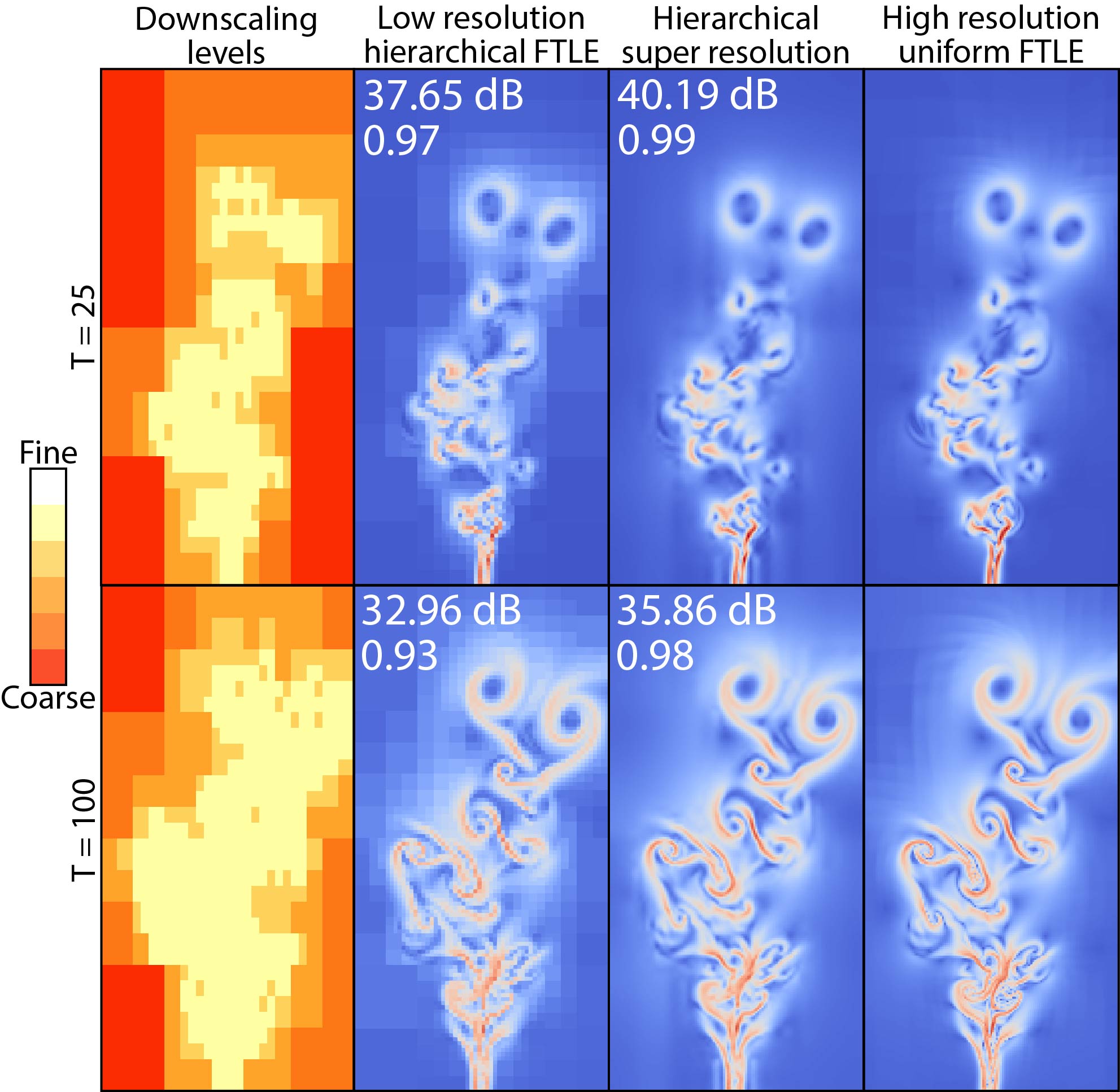}
\centering
\caption{Upscaling hierarchical FTLE data.}
\label{fig:hierarchical_FTLE}
\end{figure}

\begin{figure}[ht]
\centering
\includegraphics[width=\linewidth]{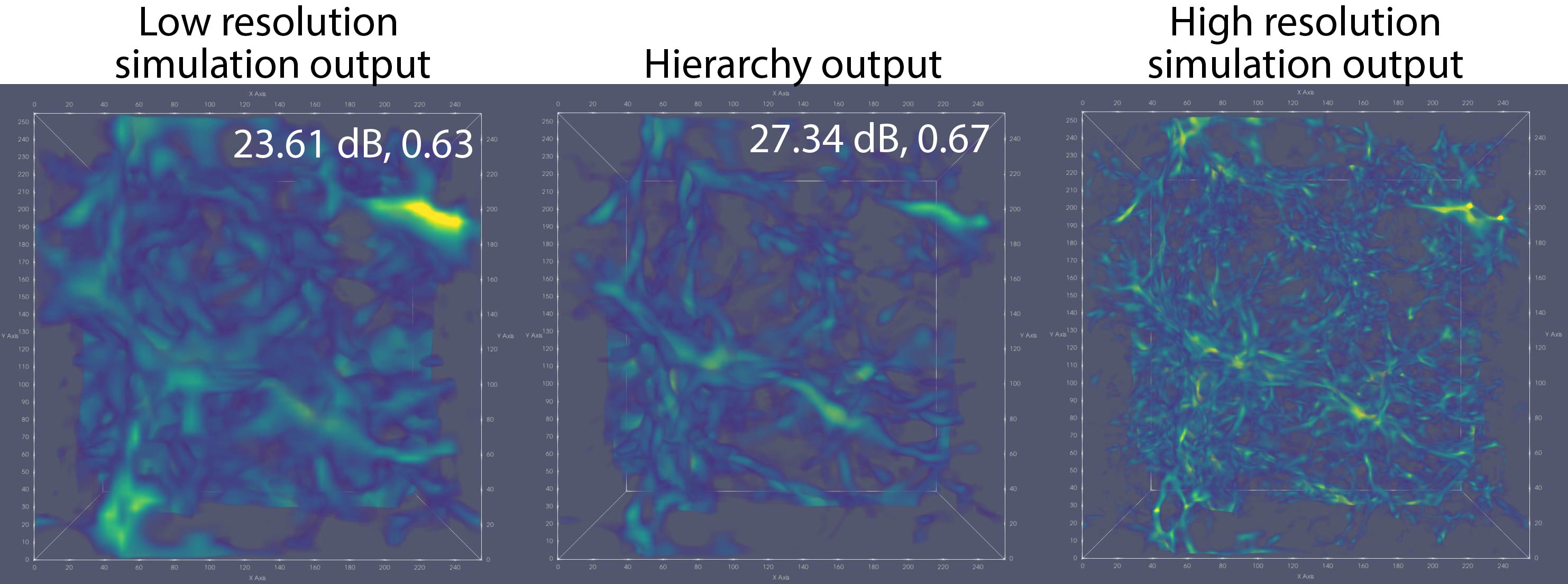}
\centering
\caption{Upscaling LR Nyx simulation output with trained neural networks for a visualization preview of what the HR simulation might look like.}
\label{fig:nyxusecase}
\end{figure}

\subsubsection{Upscaling hierarchically computed fields}
\label{hierarchical_FTLE}

\textcolor{black}{Some expensive algorithms can be computed in a hierarchical manner to adaptively allocate resources for a faster result than the same algorithm with a uniform grid.
These algorithms will often use a metric for refining the spatial domain adaptively, such as tolerable error.
Our approach can benefit these algorithms by allowing them to be run with a larger tolerable error to save more time, and performing hierarchical SR on the result for a high resolution approximation.}

\textcolor{black}{One such algorithm is FTLE, which is a useful yet expensive computation for fluid dynamics researchers to visualize Lagrangian coherent structures in flow data.
The FTLE field is computed by
\begin{equation}
    \sigma_{t_0}^T = \frac{1}{T}\ln \left( \lambda_{\text{max}}\left(\frac{\partial \phi_{t_0}^T(x)}{\partial x}'\cdot\frac{\partial \phi_{t_0}^T(x)}{\partial x}\right)\right),
\end{equation}
where $x$ is a position, $\phi$ is the flow map for the fluid data that maps a particle at position $x$ at start time $t_0$ to a position $\phi_{t_0}^{T}(x)$ after integration length $T$, and $\lambda_{\text{max}}(\cdot)$ is the maximum eigenvalue for the matrix it operates on.
Given time-varying vector field data, creating a flow map $\phi$ can be expensive, and so the FTLE field calculation can also be expensive. 
In addition, it can be difficult to know what integration length $T$ will reveal interesting features about the flow data.}

\textcolor{black}{To accelerate FTLE calculation, researchers have created hierarchical methods that adaptively sample the domain to focus on FTLE ridgelines \cite{sadlo07_AMRFTLE} or based on viewing direction \cite{barakat12_adaptiveFTLE}. 
We demonstrate that the FTLE calculation can be performed even quicker by making the subdivision requirement less strict and then using our hierarchical SR algorithm to approximate the HR result.
When an approximated HR result has features of interest, the scientist may choose to expend computing resources and time to get the full resolution result for analysis.}

\textcolor{black}{In our test, we use a dataset from a simulation with a heated cylinder with Boussinesq approximation \cite{Guenther17, gerrisflowsolver}, which is 2001 timesteps with dimensions $150 \times 450$.
We compute at total of roughly 15,000 uniform grid HR FTLE fields over $T \in [1, 5, 10, 25, 50, 100, 250, 500]$ and all possible $t_0$ for each $T$.
The dataset is randomly divided into 50\% for training the network and 50\% for a test set.
We use the trained ESRGAN hierarchy, with qualitative results shown in \autoref{fig:uniformSRquant}.}

\textcolor{black}{We use our downscaling approach as defined in \autoref{dataFormatSection} to create downscaled hierarchical data as a test case that is agnostic to the specific hierarchical FTLE computation method.
In our experiment, we use two FTLE fields calculated with $t_0=730$, $T=25$ and $t_0=805$, $T=100$, which took $3.94$ and $17.64$ seconds to calculate on a uniform grid at full resolution, respectively.
We use settings \texttt{min\_downscaling\_level=1}, \texttt{max\_downscaling\_level=5}, \texttt{min\_chunk=2}, $\epsilon=0.0193$ to generate a LR test case for the $T=25$ example, and \texttt{min\_downscaling\_level=1}, \texttt{max\_downscaling\_level=5}, \texttt{min\_chunk=2}, $\epsilon=0.02$ for the $T=100$ example, shown in \autoref{fig:hierarchical_FTLE}. 
The LR adaptive versions with our hierarchical super resolution speed up computation of the FTLE field by $16\times$ and $9.46\times$, for an estimated FTLE computation time of $0.25$ seconds and $1.86$ seconds to calculate each timestep, for $T=25$ and $T=100$, respectively.
The upscaled results achieve higher reconstruction accuracy than the nearest neighbor resampled LR hierarchical FTLE field data, and show a strong resemblance to the ground truth HR FTLE data.}

\subsubsection{Visualizing upscaled simulation results}
\label{nyx_upscaling}
\textcolor{black}{Scientists run simulations of physical phenomena which may be analyzed and visualized for a deeper understanding of the science.
The simulations may run on a fixed regular grid, in which case higher resolution simulations will take longer to finish computing than lower resolution simulations.
To save computation resources, running a lower-resolution simulation may be preferable in order to get insight into the output, before spending computation resources on the higher resolution simulation.
Super resolution methods can assist here by upscaling the low-resolution simulation output, for a visualization of what the high-resolution simulation may look like, \textcolor{black}{as shown in state of the art super resolution work for scientific data \cite{SSRVFD, SSRTVD, fukami_spatial}}.}

\textcolor{black}{To evaluate our work with this use case, we use the Nyx cosmological simulation \cite{nyx}, which uses input parameters (defined in \autoref{datasets}) to generate volumes representing the log density of dark matter.
Running the simulation at $256^3$ for a single parameter setting takes roughly 65 minutes, whereas running at $64^3$ takes only 110 seconds.
Not all parameter settings may be of interest, so to avoid unnecessary computation, we use super resolution to visualize upscaled low-resolution simulation results of certain parameters before choosing to expend computation resources on running the high resolution simulation.}

\textcolor{black}{We run Nyx with 30 parameters settings sampled randomly within the ranges listed in \autoref{datasets} at both $256^3$ and $64^3$ to generate 30 ensemble members at both low- and high-resolution.
Of those, 10 are randomly sampled for training the ESRGAN hierarchy and the other 20 for testing.}

\textcolor{black}{Over the 20 test cases, the model scored an average 26.47 dB PSNR and 0.65 SSIM, while trilinear interpolation scored an average 23.64 dB PSNR and 0.62 SSIM. 
The hierarchy also achieves a lower maximum relative error (MRE) of 0.47 compared to trilinear interpolation's result of 0.57.
\autoref{fig:nyxusecase} displays volume rendered images of LR simulation output, the neural network's approximated HR output, and the HR simulation output for parameter settings $OmM=0.14166$, $OmB=0.02288$, $h=0.61201$.
The model achieves better PSNR and SSIM than low resolution output on the example, and the volume rendering image resembles the HR output closer than the low-resolution simulation output. 
Results may improve if more than only 10 simulation output pairs are used to train the models. 
Alternatively, GAN training may improve the upscaled result's visualization to more closely resemble the high resolution output's tendril-like features at the price of lower PSNR.}
\textcolor{black}{Although this experiment does not use our hierarchical SR algorithm, the hierarchy of neural networks can still be useful. 
The simulation can be ran at any supported scale factor ($2\times, 4\times, ...$), allowing for a scalable time vs. accuracy trade-off.}

\section{Conclusion, limitations, and future work}

In this paper, we present a hierarchical SR algorithm for upscaling hierarchical data using NNs while minimizing seam artifacts on octree node boundaries. 
The hierarchical SR algorithm is comprised of a downscaling process followed by an upscaling process to reduce the effect of seam artifacts between nodes. 
We also present a hierarchy of SR NNs that can be used with our hierarchical SR algorithm to improve SR performance over bilinear/trilinear interpolation methods. 
We show that hierarchical SR has benefits over uniform SR, especially when there are coarse regions inside the volume. 
\textcolor{black}{Over three use cases, we show how our hierarchical SR approach can benefit scientific computing.}

One limitation of our method is that using $2\times$ SR NNs limits data shape, since we can only perform SR with a factor of $2^i$, for $1 \leq i \leq |G|$. 
Therefore, data can only be downscaled (and upscaled) by up to the largest factor of two of a spatial dimension, which can limit data use. 
Resampling the data to a compatible size is a short term solution (as we did for the plume dataset), but to support arbitrary spatial dimension sizes, other kinds of NN SR architecture are needed.
\textcolor{black}{Another limitation is that using multiple networks in our NN hierarchies will increase the storage overhead of the saved networks.
To reduce this overhead, a single network could be trained to perform super resolution for any scale, but it may not perform as well as the hierarchy.}
\textcolor{black}{Lastly, we want to recognize that super resolution techniques may introduce errors and artifacts, which may be critical for specific use cases.
We attempt to mitigate these drawbacks by using an error-bounded SR-octree and using ample training data so the network has learned how to upscale low-resolution features to the correct high-resolution feature.
We observe our results tend to smooth out high-frequency features as opposed to generating false features.}

In the future, our method could be improved to support non-factor-of-two volumes by using a different NN hierarchy structure, or by using continuous SR through models like LIIF \cite{LIIF}.
Our approach may also see benefits by adapting to a more general hierarchical representation such as a k-d tree instead of an octree.
\textcolor{black}{Lastly, our approach may be more attractive in the data-reduction setting with additional research toward error-bounded super resolution models, similar to current state of the art compressors \cite{TTHRESH, SZ4}, or physics informed neural networks \cite{raissi17_physicsI}.}

\ifCLASSOPTIONcompsoc
  \section*{Acknowledgments}
\else
  \section*{Acknowledgment}
\fi

This work is supported in part by the US Department of Energy SciDAC program DE-SC0021360, National Science Foundation Division of Information and Intelligent Systems IIS-1955764, and National Science Foundation Office of Advanced Cyberinfrastructure OAC-2112606. 
This work is also supported by Advanced Scientific Computing Research, Office of Science, U.S. Department of Energy, under Contract DE-AC02-06CH11357, program manager Margaret Lentz.

\ifCLASSOPTIONcaptionsoff
  \newpage
\fi



\bibliographystyle{IEEEtran}
\bibliography{IEEEabrv,manuscript.bib}
%

%

%
\begin{IEEEbiography}[{\includegraphics[width=1in,height=1.25in,clip,keepaspectratio]{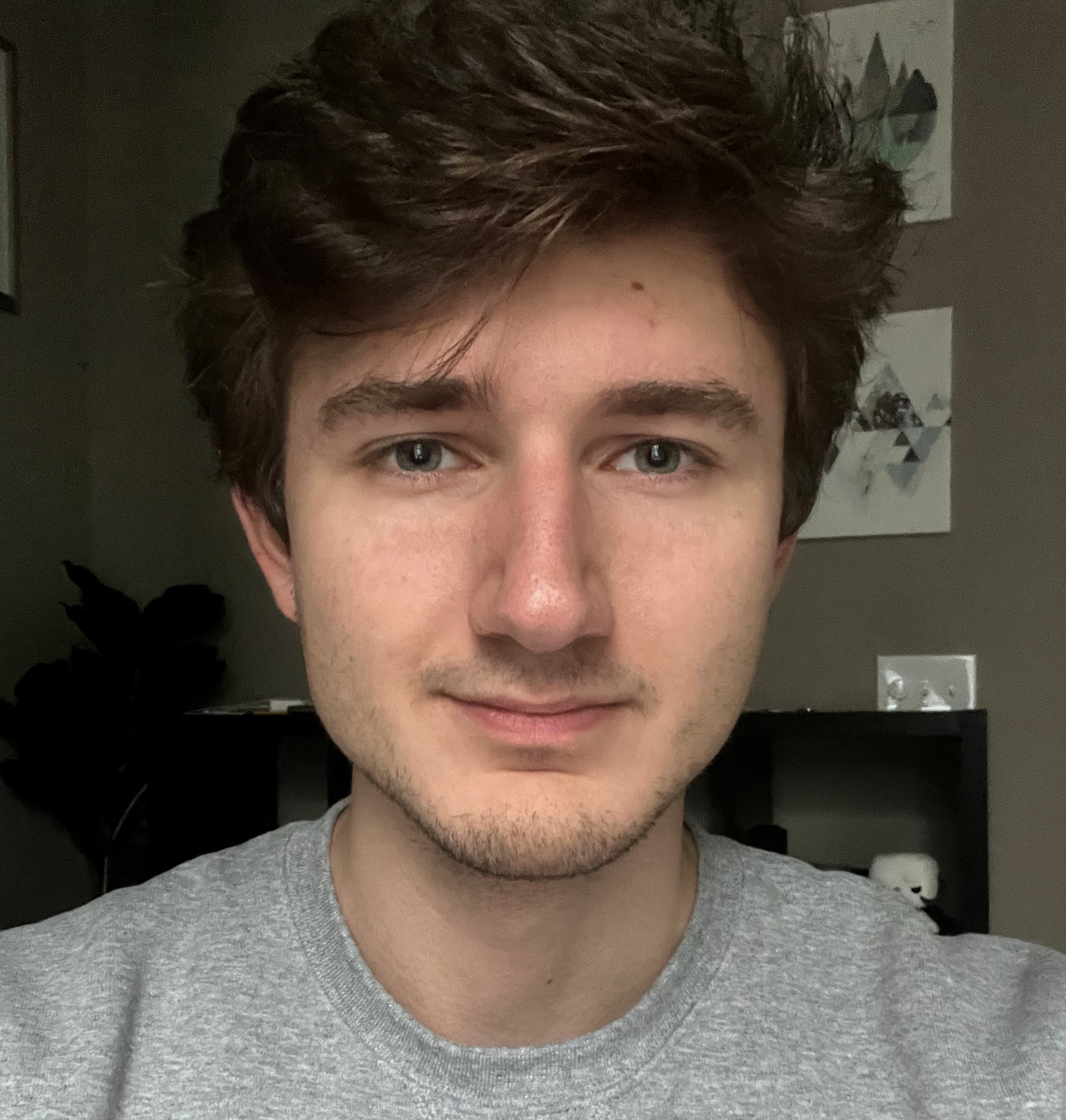}}]{Skylar W. Wurster}

is a 4th year Ph.D. student studying under Professor Han-Wei Shen as part of his GRAVITY research group at The Ohio State University in Columbus, Ohio. He also collaborates with mentors Hanqi Guo at Ohio State University and Tom Peterka at Argonne National Lab in Lemont, Illinois. His research interests span deep learning, scientific data visualization, and computer games/graphics.
\end{IEEEbiography}

\begin{IEEEbiography}[{\includegraphics[width=1in,height=1.25in,clip,keepaspectratio]{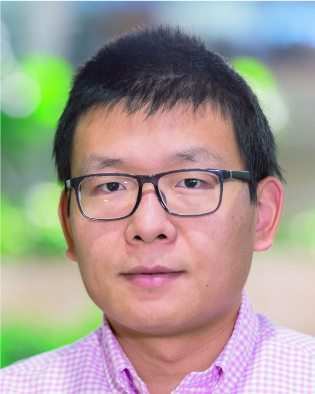}}]{Hanqi Guo} received the BS degree in mathematics and applied mathematics from the Beijing University of Posts and Telecommunications in 2009 and the PhD degree in computer science from Peking University in 2014.  He is an Associate Professor at the Department of Computer Science and Engineering in the Ohio State University.  His research interests include data analysis, visualization, and machine learning for scientific data.  He is an awardee of the DOE Early Career Research Program (ECRP) in 2022 and received multiple best paper awards in premiere visualization conferences.
\end{IEEEbiography}

\begin{IEEEbiography}[{\includegraphics[width=1in,height=1.25in,clip,keepaspectratio]{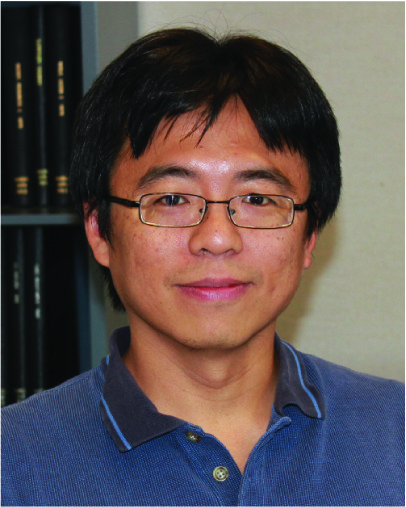}}]{Han-Wei Shen}
is a full professor at the Ohio State University. He received his B.S. degree from the Department of Computer Science and Information Engineering at National Taiwan University in 1988, his M.S. degree in computer science from the State University of New York at Stony Brook in 1992, and his Ph.D. degree in computer science from the University of Utah in 1998. From 1996 to 1999, he was a research scientist at NASA Ames Research Center in Mountain View California. His primary research interests are scientific visualization and computer graphics.  He is a winner of the National Science Foundation’s CAREER award and U.S. Department of Energy’s Early Career Principal Investigator Award. He also won the Outstanding Teaching award twice in the Department of Computer Science and Engineering at the Ohio State University.
\end{IEEEbiography}

\begin{IEEEbiography}[{\includegraphics[width=1in,height=1.25in,clip,keepaspectratio]{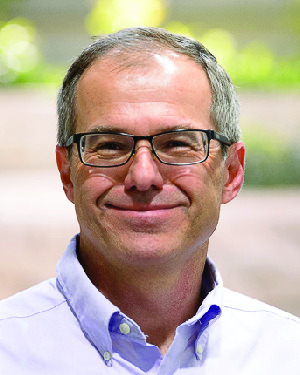}}]{Tom Peterka}
is a computer scientist at Argonne National Laboratory, scientist at the University of Chicago Consortium for Advanced Science and Engineering (CASE), and fellow of the Northwestern Argonne Institute for Science and Engineering (NAISE). His research interests are large-scale parallel in situ analysis of scientific data. Recipient of the 2017 DOE early career award and five best paper awards, Peterka has published over 100 peer-reviewed articles and papers since earning his Ph.D. in computer science from the University of Illinois at Chicago in 2007.
\end{IEEEbiography}

\begin{IEEEbiography}[{\includegraphics[width=1in,height=1.25in,clip,keepaspectratio]{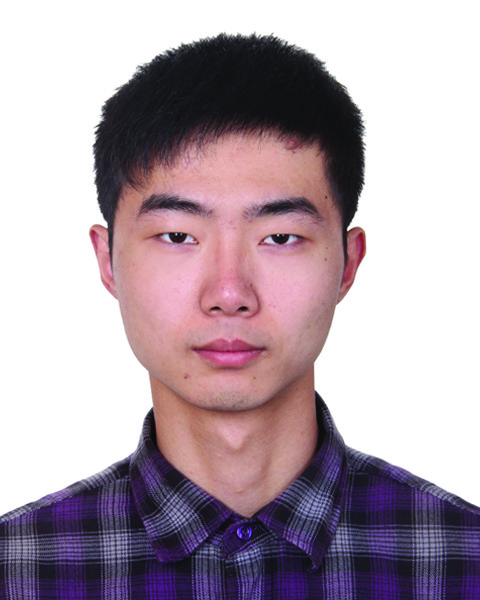}}]{Jiayi Xu}
is a research scientist at Meta AI. His research interests include high-performance data analysis, visualization, and machine learning. He is the recipient of the best paper award in the 14th IEEE Pacific Visualization Symposium. He received his Ph.D. degree in computer science and engineering from The Ohio State University in 2021 and his B.E. degree in computer science and technology from Chu Kochen Honors College of Zhejiang University in 2014.
\end{IEEEbiography}







\end{document}